\documentclass[aps,prl,twocolumn,notitlepage,longbibliography,nofootinbib,superscriptaddress,preprintnumbers]{revtex4-1}
\usepackage[utf8]{inputenc}

\usepackage{bm}
\usepackage{comment} 
\usepackage[%dvipdfmx,%<-投稿時に除く
colorlinks=true,urlcolor=blue,anchorcolor=blue,citecolor=blue,linkcolor=blue,pagecolor=blue]{hyperref}
\usepackage{amsmath,latexsym,amssymb,mathrsfs}
\usepackage{graphicx}

\usepackage{braket}
\usepackage{float}
\usepackage{mathrsfs}
\usepackage{booktabs}
\usepackage{makeidx}
\usepackage{color}
\usepackage{comment}

\usepackage{braket}

 \def\frac#1#2{{#1\over #2}}

 \def\s{\sqrt}

\def\be{\begin{equation}}
\def\ee{\end{equation}}
\def\ba{\begin{eqnarray}}
\def\ea{\end{eqnarray}}
\def\Tr{\text{Tr}}

 \def\f {\frac}

\begin{document}
\title{Late Time Quantum Chaos of pure states in the SYK model}
\author{Tokiro Numasawa}
\email{tokiro.numasawa@physics.mcgill.ca}
\affiliation{Department of Physics, McGill University, Montr\'eal, Qu\'{e}bec, H3A 2T8, Canada}
\affiliation{Department of Physics, Graduate School of Science,
Osaka university, Toyonaka 560-0043, Japan}
\preprint{OU-HET 992}
\begin{abstract}
In this letter, we study the return amplitude, which is the overlap between the initial state and the time evolved state, in the Sachdev-Ye-Kitaev (SYK) model.
Initial states are taken to be product states in a spin basis.
%Under the time evolution in the SYK Hamiltonian, the return amplitude shows the linear growth that is a signature of random matrix bahavior.
We numerically study the return amplitude by exactly diagonalizing the Hamiltonian.
We also derive the analytic expression for the return amplitude in random matrix theory.
The SYK results agree with the random matrix expectation.
We also study the time evolution under the different Hamiltonian that describes the traversable wormholes in projected black holes in the context of holography.
%The time evolution depends on the choice of product states and described by the random matrix theory
The time evolution now depends on the choice of initial product states.
The results are again explained by random matrix theory.
%Their behavior matches with the perturbation theory of quantum mechanics and random matrix theory. 
In the symplectic ensemble cases, we observed an interesting pattern of the return amplitude where  they show the second dip, ramp and plateau like behavior. 
\end{abstract}
\
\maketitle

%\section{Introduction}
{\bf{1.Introduction}}
%Since Hawking have discovered Hawking radiation, the black hole information paradox is an important but yes solved problem in gravity.
%One of these problems is that it is difficult to see the discreteness of the black hole microstate from gravity side.
The Sachdev-Ye-Kitaev (SYK) model\cite{PhysRevLett.70.3339,KitaevTalk} is an interesting model.
This model
is solvable at large $N$\cite{Maldacena:2016hyu,Kitaev:2017awl} but maximally chaotic\cite{KitaevTalk,Maldacena:2015waa}, shows random matrix behaviors \cite{PhysRevB.95.115150,Cotler:2016fpe},
and shares the same sector with two dimensional dilaton gravity\cite{JACKIW1985343,Teitelboim:1983ux,Almheiri:2014cka,Maldacena:2016upp}. 

%Recently, it is observed that the late time physics in the SYK model is governed by random matrices. 

In this letter, we study the time dependence of pure states in the SYK model.
Here we consider a special class of pure states that were first considered in \cite{Kourkoulou:2017zaj,Krishnan:2017txw} and studied further in \cite{Hunter-Jones:2017raw,Dhar:2018pii,Bhattacharya:2018fkq}.
They are simultaneous eigenstates of spin operators that are constructed from Majorana fermions.
One interesting physical interpretation of these states is that they are 
%interpreted as
states after projection measurements of maximally entangled states by these spin operators\cite{Kourkoulou:2017zaj,Numasawa:2016emc}.
In this context, we can interpret our setup as time evolution after projection measurements.
We expect that the time evolution starts flipping these spins under the SYK Hamiltonian and we get a more general superpositions of these product states.
In this paper, we consider the return amplitude, which is the square of 
the fidelity and used in the similar setup in conformal field theory\cite{Cardy:2014rqa,Cardy:2016lei}.
%the overlap between the initial state and the time evolved state. 
We can also consider the time evolution under deformed Hamiltonians after the measurements.
Here we consider a deformation proposed by \cite{Kourkoulou:2017zaj}.
They are interpreted as a traversable wormhole protocol to see the inside of black holes\cite{Kourkoulou:2017zaj,Brustein:2018fkr}.
%with in large $N$ limit and suitable choice of parameters 
This deformed Hamiltonian can also be seen as deformation of an integrable Hamiltonian with degenerate spectrum by the chaotic SYK Hamiltonian.

%\section{the SYK model}
{\bf{2.The SYK Model}}
In the SYK model, we consider even $N$ Majorana fermions $\psi_i$ that are obeyed to the anti commutation relation $\{ \psi_i , \psi_j\} = \delta_{ij}$.
The Hamiltonian of the SYK model with $q$ body interactions is given by
\be
H_{SYK} = i^{\f{q}{2}} \sum_{a_1<a_2<\cdots<a_q}  J_{a_1 \cdots a_q} \psi_{a_1}\cdots\psi_{a_q}. \label{eq:SYKHamiltonian}
\ee
Here $J_{a_1\cdots a_q}$ are random couplings with mean  $\braket{J_{a_1 \cdots a_q}}_J = 0$ and variance $\braket{J_{a_1 \cdots a_q}^2}_J = \f{J^2 (q-1)!}{N^{q-1}}$.
The $q = 4$ model is the original SYK model that we mainly focus on in this letter.
This system has two important symmetries\cite{PhysRevB.95.115150}\cite{PhysRevB.83.075103}.
The first one is the anti unitary symmetry $\mathcal{T}$.
%We can realize the fermions as the tensor products of the Pauli matrices as $\psi_{2k-1} = \f{1}{\s{2}}Z_1 \cdots Z_{k-1}X_{k}I_{k+1} \cdots I_{N/2}$ and $\psi_{2k} = \f{1}{\s{2}}Z_1 \cdots Z_{k-1}Y_{k}I_{k+1} \cdots I_{N/2}$.
%\ba
%\psi_{2k-1} &=& \f{1}{\s{2}}Z_1 \cdots Z_{k-1}X_{k}I_{k+1} \cdots I_{N/2}, \notag \\
%\psi_{2k} &=& \f{1}{\s{2}}Z_1 \cdots Z_{k-1}Y_{k}I_{k+1} \cdots I_{N/2}
%\ea
%Here $X_i,Y_i,Z_i$ are the Pauli operators on $i$-th site  and we omit the symbols $\otimes$ for tensor product.
%Then, for $N/2$ even we write $\mathcal{T} = 2^\f{N}{4}K \psi_1 \psi_3 \cdots \psi_{N-1}$ and for $N/2$ odd $\mathcal{T} = 2^\f{N}{4}K \psi_2 \psi_4\cdots \psi_{N} $. Here $K$ is the antiunitary operator that takes the complex conjugate.
This symmetry satisfies $\mathcal{T}\psi_{a}\mathcal{T}^{-1} = \psi_{a}$.
The SYK Hamiltonian (\ref{eq:SYKHamiltonian}) is invariant under $\mathcal{T}$ when $q=0\ (\text{mod}\ 4 )$.
The other important symmetry is the mod $2$ fermion number operator $(-1)^F$ with $((-1)^F)^2 = 1$.
%This is realized as $(-1)^F = 2^{\f{N}{2}}i^{-\f{N}{2}} \psi_1 \psi_2 \cdots \psi_N$.
These symmetries can have  global anomalies depending on $N$ (mod $8$)\cite{PhysRevB.83.075103}. 
These anomalies are the origin to realize all of Gaussian unitary (GUE), orthogonal (GOE) and symplectic (GSE) ensembles in the SYK model\cite{PhysRevB.95.115150}.
Here we summarize the results\cite{PhysRevB.95.115150,Kanazawa:2017dpd} in the table.\ref{table:symmetry}.
%The mixed anomaly between $\mathcal{T}$ and $(-1)^F$, which is anti commutation relation $\{\mathcal{T},(-1)^F \} = 0$ rather than commutation relation $[\mathcal{T},(-1)^F] = 0$, exists in $N=2,6\ (\text{mod}\ 8)$ cases.
%In this case $\mathcal{T}$ exchanges the value of $(-1)^F$.
%The antiunitary symmetry $\mathcal{T}$ does not put any restriction on each block with fixed $(-1)^F$ and we expect GUE statistics.
%The eigenvalues of each block are related by anti commutation relation and we obtain exact $2$ degeneracy in each energy level. 

%..
%The time reversal anomaly, which means $\mathcal{T}^2 = -1$ rather than $\mathcal{T}^2=1$, level statistics becomes GSE with exact 2 degeneracy that is so called Kramers degeneracy. 
%This anomaly occurs in $N=4,6\ (\text{mod}\ 8)$, but in $N=6$ case we already have the mixed anomaly with $(-1)^F$ and the statistics are GUE.
%Therefore, we expect GOE statistics for $N=0\ (\text{mod}\ 8 )$, GUE statistics for $N=2,6\ (\text{mod}\ 8)$ and GSE statistics for $N=4\ (\text{mod}\ 8)$
\begin{table}[htb]
  \begin{center}
    \caption{symmetry property in the SYK model}\label{table:symmetry}
    \begin{tabular}{|c||c|c|c|c|} \hline
      $N\ (\text{mod}\ 8)$  & $\mathcal{T}^2$ & $\mathcal{T}(-1)^F = a (-1)^F\mathcal{T}$ & statistics &degeneracy\\  \hline
      $N=0$ & $+1$ & $a=+1$ & GOE &1 \\
      $N=2$ & $+1$ & $a=-1$ & GUE &2 \\
     $N=4$ &  $-1$  & $a=+1$ & GSE&2 \\
     $N=6$ &  $-1$  & $a=-1$  & GUE &2 \\ \hline
    \end{tabular}
  \end{center}
\end{table}

Now, we consider the return amplitude.
The return amplitude is 
\be
g_p(t) = |\bra{\psi_0}e^{- i Ht}\ket{\psi_0}|^2. \label{eq:RAdef}
\ee
Here $\ket{\psi_0}$ is a initial state and $H$ is the Hamiltonian of the system.
This definition is also applicable to any quantum systems.
We take the square of $|\bra{\psi_0}e^{- i Ht}\ket{\psi_0}|$, which is different from the definition in \cite{Cardy:2014rqa,Cardy:2016lei}.
Our choice makes the relation to so called spectral form factor clear, which is studied extensively in the field of quantum chaos and also recently in holography\cite{Cotler:2016fpe,Balasubramanian:2016ids,Benjamin:2018kre,Chenu:2018spm,delCampo:2017bzr}.
%as we will see later

%(これでいい？)
Because we expect that the late time behavior of the SYK model is governed by random matrices \cite{Cotler:2016fpe}, it is good to study first the return amplitude in random matrix theory.
%To calculate (\ref{eq:RAdef}) in random matrix theory, 
%It is convenient to rewrite (\ref{eq:RAdef}) by using the projection operator $\Pi = \ket{\psi_0} \bra{\psi_0}$ on to the one dimensional subspace that is spanned by $\ket{\psi_0}$.
%Then, we can write (\ref{eq:RAdef}) as the two point function of projection operators:
%\be
%g_p(t) =\Tr(\Pi(t) \Pi(0)) = \Tr( e^{-iHt} \Pi e^{iHt}\Pi).
%\ee
%Therefore, the return amplitude is expressed as a two point function of the projection operator on a thermal background with infinite temperature.
Now we want to compute $ \braket{|\bra{\psi_0}e^{- i Ht}\ket{\psi_0}|^2}_{\text{GUE}} = \int dH e^{-\f{L}{2}\Tr H^2} |\bra{\psi_0}e^{- i Ht}\ket{\psi_0}|^2$ where $dH$ is the Haar measure on the space of $L\times L$ Hermitian matrices. 
We can compute this in random matrix theory using the Haar integrals.
%The measure  is invariant under unitary conjugation $d(U^{\dagger} H U) e^{-\f{L}{2}\Tr (U^{\dagger}HU)^2} = dH e^{-\f{L}{2}\Tr H^2}$, we can represent the GUE ensemble average as $ \braket{f(H)}_{\text{GUE}} = \int dH \int dU e^{-\f{L}{2}\Tr H^2} f(U^{\dagger}H U)$ for any function $f(H)$ where $dU$ is the Haar measure.
%By taking the $dU$ integral first, we obtain 
The results are
\be
\braket{|\bra{\psi_0}e^{-iHt}\ket{\psi_0}|^2}_{\text{GUE}} = \f{1}{L(L+1)} (\braket{g(t)}_{\text{GUE}} + L), \label{eq:RAandSFF}
\ee
where $g(t) =  Z(t)Z(t)^*$ with $Z(t) = \Tr( e^{- i H t})$.
This $g(t)$ is the spectral form factor, which diagnoses the energy level correlations in chaotic systems.
We also find that this relation also holds in GSE ensemble by replacing the aevrage to $\braket{}_{GSE}$, using the Haar integral\cite{2006CMaPh.264..773C,1997PhRvB..55.1142A}.
In GOE ensemble, the return amplitude depends on the initial state $\ket{\psi_0}$\footnote{The return amplitude in GOE ensemble is also studied in \cite{2018arXiv180707577S}.}. 
The equation (\ref{eq:RAandSFF}) means that the return amplitude is given essentially by the constant $L$ shift of the spectral form factor.
%Another interesting point of (\ref{eq:RAandSFF})  is that the return amplitude does not depend on the initial state in random matrix theory.
When $t=0$, the spectral form factor is simply given by the square of the dimension of the Hilbert space $g(0) =L^2$.
Under the time evolution, $\braket{g(t)}_{\text{GUE}}$ decreases and hit the minimal value. These regimes are called the slope and the dip \cite{Cotler:2016fpe}.
Then, $\braket{g(t)}_{\text{GUE}}$ increases linearly. 
This linear growth is called the ramp\cite{Cotler:2016fpe} and this reflect the long range eigenvalue correlations in chaotic systems.
Finally $\braket{g(t)}_{\text{GUE}}$ saturate the late time value $\braket{g(\infty)}_{\text{GUE}} = L$ in sufficiently late time.
This is called the plateau\cite{Cotler:2016fpe}.
The plateau value generically coincides with the infinite time average:
\be
 \lim_{T\to \infty} \f{1}{T} \int _0^T dt \sum_{m,n} e^{-i(E_m-E_n)t} = \sum_E N_E, \label{eq:plateau}
\ee 
where $N_E$ is the degeneracy of each energy level $E$.
This plateau value is much smaller than the initial value $g(0) = L^2$ but still bigger than $\mathcal{O}(1)$.
%ate time the spectral form factor reduces to much smaller value $g(\infty) = L$.
The relation (\ref{eq:RAandSFF}) between the return amplitude $g_p(t)$ and the spectral form factor $g(t)$ says that the return amplitude also shows the slope, the dip, the ramp and the plateau.
%Therefore, under the time evolution return amplitude decreases and at late time the return amplitude reduces to $g_p(\infty) = 2/(L+1)$, which is much smaller than the initial value $g_p(0) = 1$.
%(\ref{eq:RAandSFF}) also says that, before saturating the late time value, the return amplitude also shows the linear growth that is a signal of eigenvalue correlations in chaotic systems.
The plateau value for the return amplitude is given by  $\braket{g_p(t)}_{\text{GUE}} = 2/(L+1)$, which is also much smaller than the initial value $g_p(0) = 1$.

(\ref{eq:RAandSFF}) also means that the initial state can $\ket{\psi_0}$ evolves to other states that are orthogonal to $\ket{\psi_0}$.
Let us pick a state $\ket{\psi_1}$ that satisfies $\braket{\psi_1|\psi_0} = 0$.
In the similar way with the return amplitude, we can calculate the overlap $ |\bra{\psi_1}e^{-iHt} \ket{\psi_0}|^2$.
That becomes 
\be
\braket{|\bra{\psi_1}e^{-iHt} \ket{\psi_0}|^2 }_{\text{GUE}} = \f{1}{L^2-1} \Big(L - \f{\braket{g(t)}_{\text{GUE}}}{L} \Big). \label{eq:EAdef}
\ee 
This amplitude increases under the time evolution from $0$, then slightly decreases and finally saturate the late time value $1/(L+1)$.
Interestingly, this late time value is not equal to the late time value of the return amplitude $g_p(\infty) = 2/(L+1)$ but the half of that.
On the other hand, we get $\int dU |\bra{\psi} U \ket{\phi}|^2 = 1/L$ for any choice of $\ket{\psi}$ and $\ket{\phi}$ for Haar random unitary $U$.
This is because the average with $\lim _{T\to \infty} \int dH e^{-\f{L}{2}\Tr H^2} F(e^{-iHt}) $ is not equivalent to the Haar random unitary average $\int dU F(U)$ for functions $F$ on the space of $L\times L$ unitary groups\cite{Saad:2018bqo}.

The spectral form factor have a finite temperature generalization $g(t;\beta) = \Tr( e^{-\beta H- i H t })\Tr (e^{-\beta H+ i H t })$.
A finite temperature analog of the return amplitude, we consider 
\be
g_p(t;\beta) = |\bra{\psi_0}e^{-\beta H - i H t }\ket{\psi_0}| ^2 .\label{eq:FTRAdef}
\ee
We can think of this as the return amplitude with the initial state $e^{-\f{\beta}{2}H}\ket{\psi_0}$.
As is the case with the finite temperature spectral form factor\cite{Cotler:2016fpe}, when we take the ensemble average
% in random matrix theory or in the SYK model
, the annealed disorder $\f{\braket{|\bra{\psi_0}e^{-\beta H - i H t }\ket{\psi_0}| ^2}_E}{\braket{|\bra{\psi_0}e^{-\beta H}\ket{\psi_0}| ^2}_E} $ is not equal to the quenched disorder $ \braket{\f{|\bra{\psi_0}e^{-\beta H - i H t }\ket{\psi_0}| ^2}{|\bra{\psi_0}e^{-\beta H}\ket{\psi_0}| ^2}}_E$ where $\braket{}_E$ is an ensemble average.
In this paper we consider the annealed disorder in which the analytic treatment becomes easy\cite{Cotler:2016fpe,Cotler:2017jue}.
Another motivation to take the annealed version is its similarity with the unnormalized cylinder amplitude in Quantum Field Theories\cite{Cardy:2014rqa,Cardy:2016lei}.
%In infinite temparature this difference does not show up because the denominator $\braket{\psi_0|\psi_0}$ is not affected by the disorder average.
%That means $\f{\braket{|\bra{\psi_0}e^{-\beta H - i H t }\ket{\psi_0}| ^2}_E}{\braket{|\bra{\psi_0}e^{-\beta H}\ket{\psi_0}| ^2}_E} \neq \braket{\f{|\bra{\psi_0}e^{-\beta H - i H t }\ket{\psi_0}| ^2}{|\bra{\psi_0}e^{-\beta H}\ket{\psi_0}| ^2}}_E$ where $\braket{}_E$ means an ensemble average.
The ensemble average of the finite temperature return amplitude (\ref{eq:FTRAdef}) in random matrix theory becomes 
\ba
&&\braket{|\bra{\psi_0}e^{-\beta H - i H t }\ket{\psi_0}| ^2}_{\text{GUE}}  \notag \\
&=& \f{1}{L(L+1)}(\braket{g(t;\beta)}_{\text{GUE}} + \braket{g(0;2\beta)}_{\text{GUE}}). \label{eq:shiftedSFF}
\ea
We call the right hand side of (\ref{eq:RAandSFF}) and (\ref{eq:shiftedSFF}) the {\it{shifted spectral form factor}}.
%To , we should include this smearing factor to excite only low energy modes.

\begin{figure}[htbp]
\begin{center}
\includegraphics[width=\hsize]{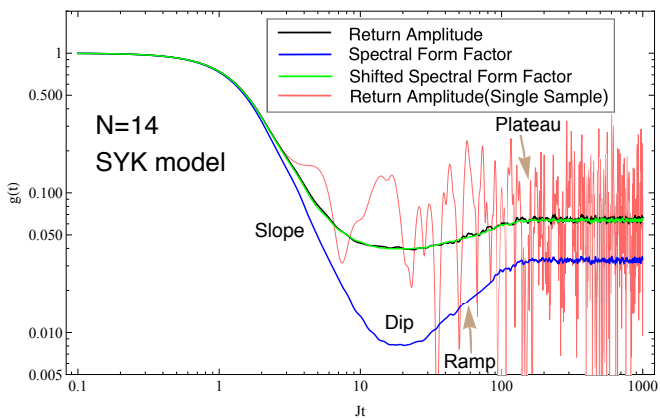}
\caption{These are numerical plots for the $N=14$ $q=4$ SYK model.
%The level statistics are given by GUE.
We take the disorder average for $1500$ samples except for the single sample case.
We put $\beta = 1.5$.
The return amplitude is defined (\ref{eq:SYKRAdef}) and we choose the state that satisfies $S_k\ket{B_{\uparrow\cdots \uparrow}} = \ket{B_{\uparrow\cdots \uparrow}}$.
The shifted spectral form factor is the right hand side of (\ref{eq:shiftedSFF}) for the SYK Hamiltonian.
We normalize them so that the initial values become $1$. 
%When we take the disorder averages, we treat them as annealed disorder that means we take first the disorder average for unnormalized quantities and them divide them by the initial value of them. 
A single sample of $g_p(t)$ shows erratic oscillation around the averaged return amplitude at late time and it is not self averaging\cite{1997PhRvL..78.2280P}.
}
\label{fig:RAN14}
\end{center}
\end{figure}

Now we consider the return amplitude in the SYK model.
The states we consider is defined as follows.
First there are $N/2$ spin operators in the SYK model defined by $S_k = - 2i \psi_{2k-1}\psi_{2k}$.
These satisfy $S_k^2 = 1$ and eigenvalues of $S_k$ are given by $s_k =\pm  1$.
The pure states $\ket{B_s}$ we consider is defined as common eigenstates of these spin operators:
\be
S_k\ket{B_s} = s_k \ket{B_s}, \qquad \text{for}\ k = 1 ,\cdots , N/2. 
\ee
This defines $2^{\f{N}{2}}$ states, one for each choice of the spins $s_k$.
These states form a basis of the SYK Hilbert spaces.
By including the Euclidean evolution in the SYK Hamiltonian, we can produce lower energy states:
\be
\ket{B_s(\beta)} = e^{-\f{\beta}{2}H_{SYK}} \ket{B_s}.
\ee
The return amplitude for the SYK model are 
\be
g_p(t;\beta) =|\bra{B_s}e^{- \beta H_{SYK}-iH_{SYK}t}\ket{B_s}|^2. \label{eq:SYKRAdef}
\ee
Currently we do not have any technique to compute analytically (\ref{eq:SYKRAdef}) for finite $N$,
we numerically computed the finite temperature return amplitude (\ref{eq:SYKRAdef}) in the SYK model.
The results are plotted in Fig\ref{fig:RAN14}.
Clearly, we observe the slope, the dip, the ramp and the plateau in the return amplitude in the SYK model.
The early time decay is almost the same with the spectral form factor.
%The spectral form factor is the product $\Tr (e^{-\beta H -i Ht })\Tr (e^{-\beta H +i Ht })$, but the disorder average for $j_{a_1\cdots a_q}$ introduce the interaction term between the degrees of freedom for $\Tr (e^{-\beta H -i Ht })$  and $\Tr (e^{-\beta H +i Ht })$ .
In the large $N$ limit, $\bra{B_s}e^{-\beta H_{SYK}} \ket{B_s} = 2^{-\f{N}{2}}\Tr( e^{-\beta H_{SYK}}) + \mathcal{O}(1/N^{q-1})$ for any $\ket{B_s}$ \cite{Kourkoulou:2017zaj} in the leading of $1/N$ expansion.
The early time dependence is captured by the analytic continuation of $\beta \to \beta + it$ from the leading term in $1/N$, we expect the match between them.
On the other hand, the ramp and the plateau region we do not expect that because they are non perturbative effects in $1/N$ expansion\cite{Cotler:2016fpe,Saad:2018bqo}.
The plot shows that they take different value at late time.
% this is expected because the replica diagonal part is dominant in $1/N^{q-1}$ expansion.
The late time behavior is expected to be governed by random matrix theory\cite{Cotler:2016fpe}.
We also expect this to the return amplitude.
To confirm this, we compare the return amplitude with the shifted spectral form factor (\ref{eq:shiftedSFF}) where the ensemble average $\braket{g(t;\beta)}_{\text{GUE}}$ is replaced by the SYK coupling average $\braket{g(t;\beta)}_{J}$. 
We also restrict the Hamiltonian to the fixed $(-1)^F$ charge sector in the shifted spectral form factor because only that acts on the state $\ket{B_s}$.
The plots agree very well and these results also support the random matrix behavior in the late time in the SYK model.

\begin{figure}[htbp]
\begin{center}
\includegraphics[width=\hsize]{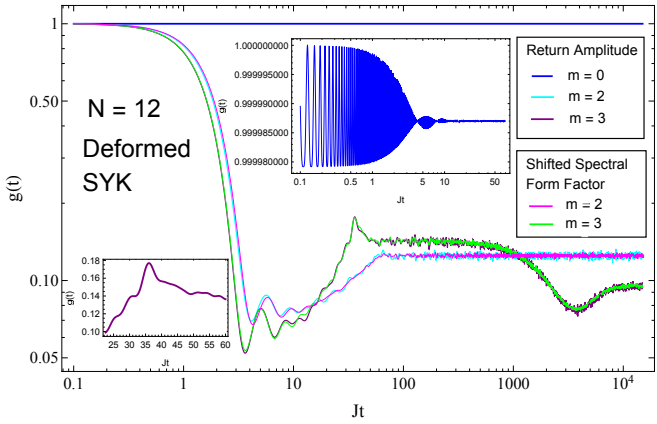}
\caption{These are numerical plots for the $N=12$ $q=4$ deformed SYK model with $\mu = 50$ and $s_k = 1$ in (\ref{eq:MassH}) for all $k$.
We take the average over $2000$ samples.
$m$ is the number of flips of spins from the B state from the ground state of $H_M$.
More explicitly, we choose $\ket{B_{\uparrow\uparrow\uparrow\uparrow\uparrow\uparrow}}$ for $m=0$, $\ket{B_{\downarrow\downarrow\uparrow\uparrow\uparrow\uparrow}}$ as an example of $m=2$ cases  and $\ket{B_{\downarrow\downarrow\downarrow\uparrow\uparrow\uparrow}}$ for $m=3$.
We also plotted the shifted spectral form factor of band $m$ where
the spectral form factor is restricted on the $m$-th band  is $\sum_{i,j = i_m}^{f_m} e^{-i(E_i-E_j)t}$
% is restricted betweem $i=d_1 + \cdots + d_{m-1}+1$ to $i  = d_1 + \cdots + d_{m-1}+d_m$
and the shift (\ref{eq:RAandSFF}) is given by $L = d_m$. 
We see the kink around the transition time from the ramp to the plateau in the $m=3$ case, which reflects GSE statistics\cite{Cotler:2016fpe}.
}
\label{fig:RAN12Def}
\end{center}
\end{figure}

%\subsection{A deformed Hamiltonian}
{\bf{3. A Deformed Hamiltonian}}
Next we consider the following "mass term" Hamiltonian\cite{Kourkoulou:2017zaj}:
\be
H_M = -\f{1}{2}\sum_{k} s_k S_k = i  \sum_{k} s_k \psi_{2k-1}\psi_{2k}. \label{eq:MassH}
\ee
This Hamiltonian is diagonalized in the $\ket{B_s}$ state basis.
Especially, the unique ground state of this Hamiltonian is given by $\ket{B_s}$ with spin $\{s_k\}$ and energy $E_0^{(0)} = -N/4$.
By flipping some spins from the ground state $\ket{B_s}$, we obtain the whole energy eigenstates.
The excited state energy levels are given by 
\be
E_m^{(0)} = -\f{N}{4} + m \  \text{with degeneracy} \  d_m =  \begin{pmatrix}
 N/2 \\ m
\end{pmatrix}. \label{eq:massHspectrum}
\ee
%\ee
%and the degeneracy of the $m$-th excited states is 
%\be
%d_m =  \begin{pmatrix}
% N/2 \\ m
%\end{pmatrix}.
%\ee
There are energy gaps , which is given by $E_{m+1}^{(0)} - E_m^{(0)}=1$, between the bands.
%Each energy band is separated by the gap $1$ from the next energy band.
Now we consider the Hamiltonian that contains the both of the SYK term and (\ref{eq:MassH}):
\be
H_{def} = H_{SYK} + \mu H_M.
\ee
This Hamiltonian was originally proposed to describe the traversable wormhole after projection measurements\cite{Kourkoulou:2017zaj}.
We call this $H_{def}$  the {\it deformed Hamiltonian}.
%Without loss of generality, we can put $s_k = 1$ for all the $k$. 
%The ground state of the mass term (\ref{eq:MassH}) is given by $\ket{B_{\uparrow \cdots \uparrow}}$.
Here we consider the regime that $\mu$ is large and we can treat the SYK term as perturbation.
This can be seen as a perturbation of the integrable system with degenerate spectrum by the chaotic Hamiltonian.
\footnote{Other kind of mass deformations are considered in \cite{Eberlein:2017wah,Garcia-Garcia:2017bkg,Nosaka:2018iat}}.
We also concentrate on the infinite temperature cases.
Because $\mu$ is large, exact energy levels from $E_{i_m}$ with $i_m = d_1 + \cdots + d_{m-1}+1$ to $E_{f_m}$  with $f_m= d_1 + \cdots + d_{m}$ localize near $E_m^{(0)}$ and form  band like structure.
By exactly diagonalizing the Hamiltonian, we can study the return amplitude under the deformed Hamiltonian.
We show the numerical results in Fig.\ref{fig:RAN12Def}.
Here we explain the results.
First, the ground state of $H_{def}$ is almost given by $\ket{B_s}$ with $m=0$ because the gap between the ground state and the first excited states is $\mu$, which is sufficiently large and suppresses the mixing with other  $\ket{B_s}$ states.
In this case, the return amplitude does not decay and also shows oscillation at early time.
%We also expect that the perturbation by $H_{SYK}$ removes the degeneracy in $H_M$.
The behavior of $\ket{B_s}$ with $m\neq 0$ are described by random matrix theory.
According to the perturbation theory of quantum mechanics the first order shift of energy levels are determined by the projection of $H_{SYK}$ onto the degenerate energy levels\cite{messiah1999quantum}.
Therefore at early time the projection of $H_{SYK}$ on each degenerate energy levels determines the time evolution of $\ket{B_s}$.
We know the dimension of the projected Hamiltonian from (\ref{eq:massHspectrum}).
By considering the spectral form factor restricted on the $m$-th band and its shift by $L = d_m$, we see the very good agreement with the return amplitude.
Because the spectral form factor and the return amplitude depend on the level statistics, next we determines the symmetry class of each band.
To see this, we need to know the symmetry properties of $\ket{B_s}$.
These are eigenstates of $(-1)^F$.
The anti unitary $\mathcal{T}$ flips the spin $\mathcal{T} S_k \mathcal{T}^{-1} = -S_k$.
Therefore, we find that $\mathcal{T}\ket{B_s} = e^{i\theta_s} \ket{B_{-s}}$ where $e^{i\theta_s}$ is a phase factor and $\ket{B_{-s}}$ is the state that satisfies $S_{k}\ket{B_{-s}} = -s_k\ket{B_{-s}}$. 
From this, we find for the projection operator onto the flip number $m$ sector $P_m = \sum_{ \#\text{flip} = m} \ket{B_s}\bra{B_s}$ satisfies 
\be
\mathcal{T}P_m\mathcal{T}^{-1} = P_{\f{N}{2}-m}.
\ee
This means that $\mathcal{T}$ flips the bands.

Now we consider the symmetry class based on these properties.
First we consider the $m\neq \f{N}{4}$ bands.
In this case, $H_m$ satisfies $\mathcal{T}H_m \mathcal{T}^{-1} = H_{\f{N}{2}-m}$.
Therefore, there are no constraint from the symmetry $\mathcal{T}$ and we expect GUE statistics for $H_m$.
%$m=\f{N}{4}$th bands in $N=0,4\ (\text{mod}\ 8)$ cases are special because they are mapped by $\mathcal{T}$ to themselves.
Next we consider $m=\f{N}{4 }$th bands which exist in $N = 0,4\ (\text{mod}\ 8)$ cases.
Now $\mathcal{T}$ imposes symmetry constraint $\mathcal{T}H_\f{N}{4}\mathcal{T}^{-1} = H_\f{N}{4}$ and we expect GOE ensemble for $N=0\ (\text{mod}\ 8)$ and GSE ensemble for $N=4\ (\text{mod}\ 8)$ on these bands at the first order of perturbation.
Except for $m=\f{N}{4}$th bands in $N=4\ (\text{mod}\ 8)$, we expect that degeneracies are completely removed by the SYK Hamiltonian perturbation.
Because of Kramers' degeneracies of $H_{\f{N}{4}}$ that originate to the time reversal anomaly $\mathcal{T}^2 = -1$, $m=\f{N}{4}$th bands in $N=4\ (\text{mod}\ 8)$ have still two degeneracies at each level at the first order perturbation. 
According to the second order perturbation theroy\cite{messiah1999quantum},
the second order shift of degenerate spectrums is determined by $H_{\f{N}{4}}^i = \mathcal{P}_i H_{SYK}\mathcal{Q}_\f{N}{4} H_{SYK}\mathcal{P}_i$.
Here $\mathcal{Q}_{\f{N}{4}} = \sum_{m\neq \f{N}{4}} \f{P_m}{-\f{N}{4}+m}$ 
and $\mathcal{P}_i = \ket{\psi_{i,1}}\bra{\psi_{i,1}} +\ket{\psi_{i,2}}\bra{\psi_{i,2}}$ with two eigenstates of $i$-th degenerate eigenvalues.
We choose the basis that satisfies $\mathcal{T}\ket{\psi_{i,1}} = \ket{\psi_{i,2}}$.
From these definitions, we find that $\mathcal{T}\mathcal{P}_i\mathcal{T}^{-1} = \mathcal{P}_i$  and $\mathcal{T}\mathcal{Q}_{\f{N}{4}}\mathcal{T}^{-1} = - \mathcal{Q}_{\f{N}{4}}$.
This means $\mathcal{T}H_{\f{N}{4}}^i \mathcal{T}^{-1} = -H_{\f{N}{4}}^i$.
By solving this symmetry constraint in the basis with
$
\mathcal{T} = 
\begin{pmatrix}
 0 & -1 \\
 1 & 0 
\end{pmatrix} K
$
, we obtain
\be
H_{\f{N}{4}}^i = \mathcal{P}_i H_{SYK}\mathcal{Q}_\f{N}{4} H_{SYK}\mathcal{P}_i=
\begin{pmatrix}
 x & a \\
 a^* & -x 
\end{pmatrix},  \label{eq:2ndorder}
\ee
with a real number $x$ and a complex number $a$.
This means that in a generic Hamiltonian the off diagonal element $a$ enters and the degeneracies are removed.
%This is important because the plateau values of the return amplitude and the spectral form factor depend on the degeneracy.
Therefore the degeneracy is removed at the second order of the perturbation.
This difference of the order means that the return amplitude (and the spectral form factor) does not see the true degeneracy at early time.
We see numerically in fig.\ref{fig:RAN12Def} that they show the first order degeneracy at the first plateau, but after that they show the second slop, dip, ramp, and plateau.
The second plateau value is smaller than the first plateau value because (\ref{eq:plateau}) means that degeneracies give larger plateau values.

To see the level statistics further, we study the distribution of the adjacent gap ratio\cite{2015PhRvB..91h1103L,2007PhRvB..75o5111O,2016PhRvB..94n4201B,PhysRevB.95.115150} for each $m$-th band.
The adjacent gap ratio is defined as $r_n = \f{E_{n+1} - E_n }{E_n - E_{n-1}}$ for an ordered spectrum $E_{n-1}< E_n < E_{n+1}$.
The distribution of the ratio $r_n$ in Poisson statistics is 
\be
p(r) = \f{1}{(1+r)^2}. \label{eq:WSpoisson}
\ee
On the other hand, in random matrices, the distribution of the ratio $r_n$ becomes\cite{2013PhRvL.110h4101A} 
\be
p(r) = \f{1}{Z_{\beta}}\f{(r+r^2)^{\beta}}{(1+r+r^2)^{1 + \f{3}{2}\beta}}.\label{eq:WSrandom}
\ee
For GOE $\beta = 1$ and $Z_{\beta} = \f{27}{8}$. For GUE $\beta = 2$ and $\f{4\pi}{81\s{3}}$. For GSE $\beta = 4$ and $Z_{\beta} = \f{4\pi}{729}$.
$r\to 0$ behavior $p(r)\sim r^{\beta}$ represents the level repulsion.
We study this gap ratio numerically in $N=16$ case and compare with the random matrix case.
The $m = \f{N}{4} =4$th band shows GOE statistics and the $m=3$rd band shows GUE statistics.
These result agree with our symmetry analysis above.
\begin{figure}[htbp]
\begin{minipage}{0.49\hsize}
\begin{center}
\includegraphics[width=4.3cm]{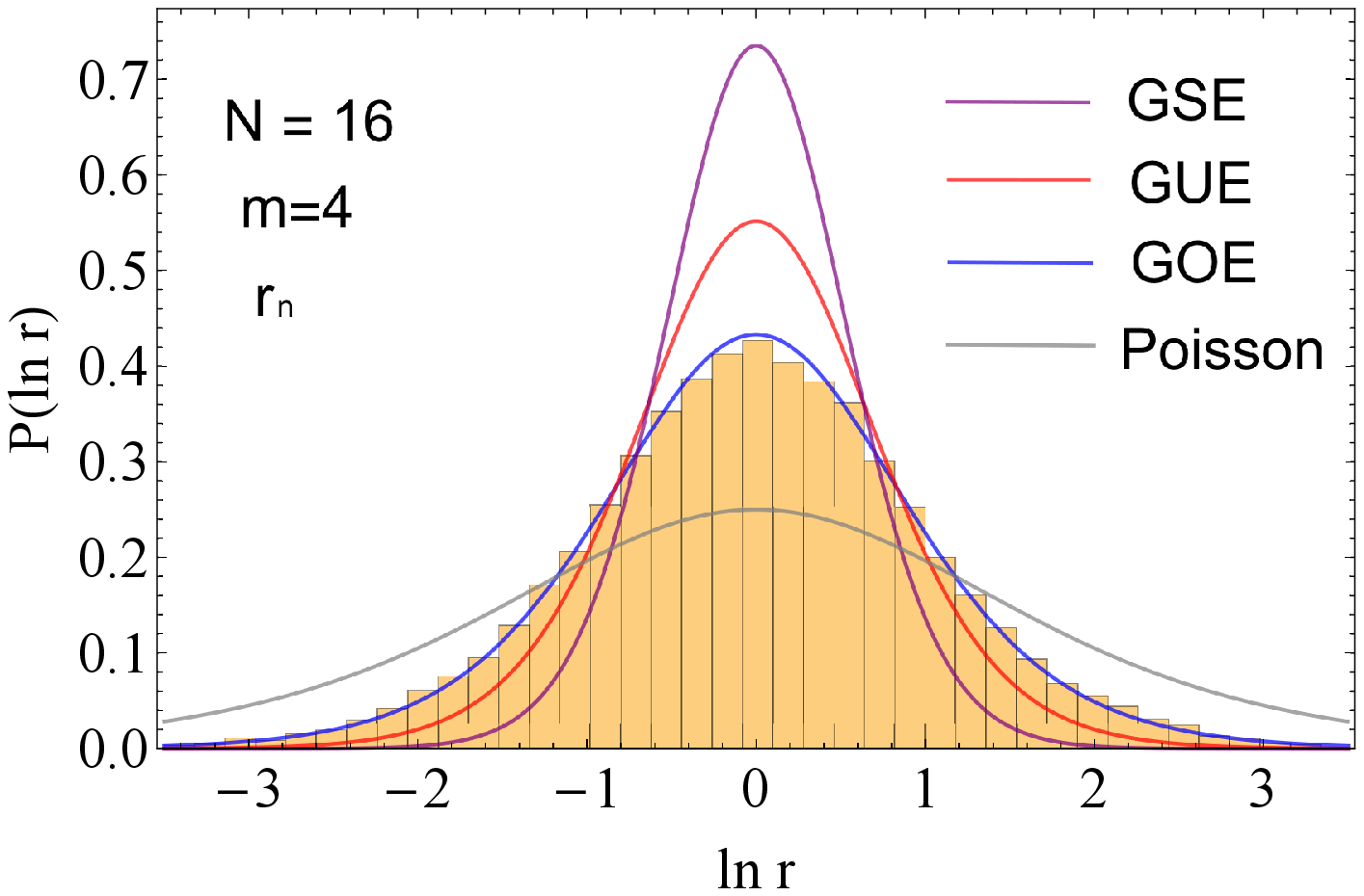}
%\subcaption{Positive tension} 
\end{center}
\end{minipage}
\begin{minipage}{0.49\hsize}
\begin{center}
\includegraphics[width=4.3cm]{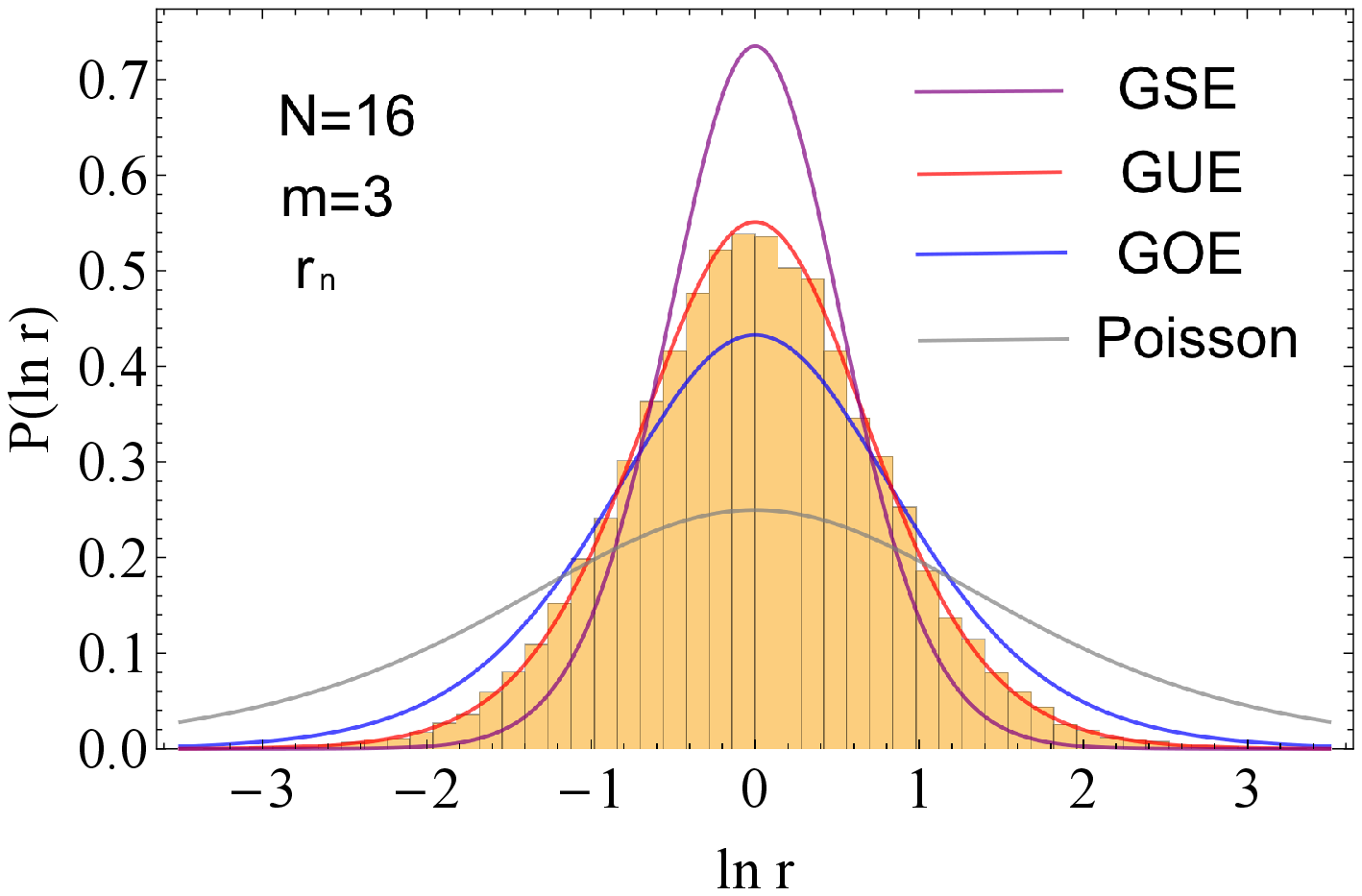}
%\subcaption{Positive tension} 
\end{center}
\end{minipage}
\begin{minipage}{0.49\hsize}
\begin{center}
\includegraphics[width=4.3cm]{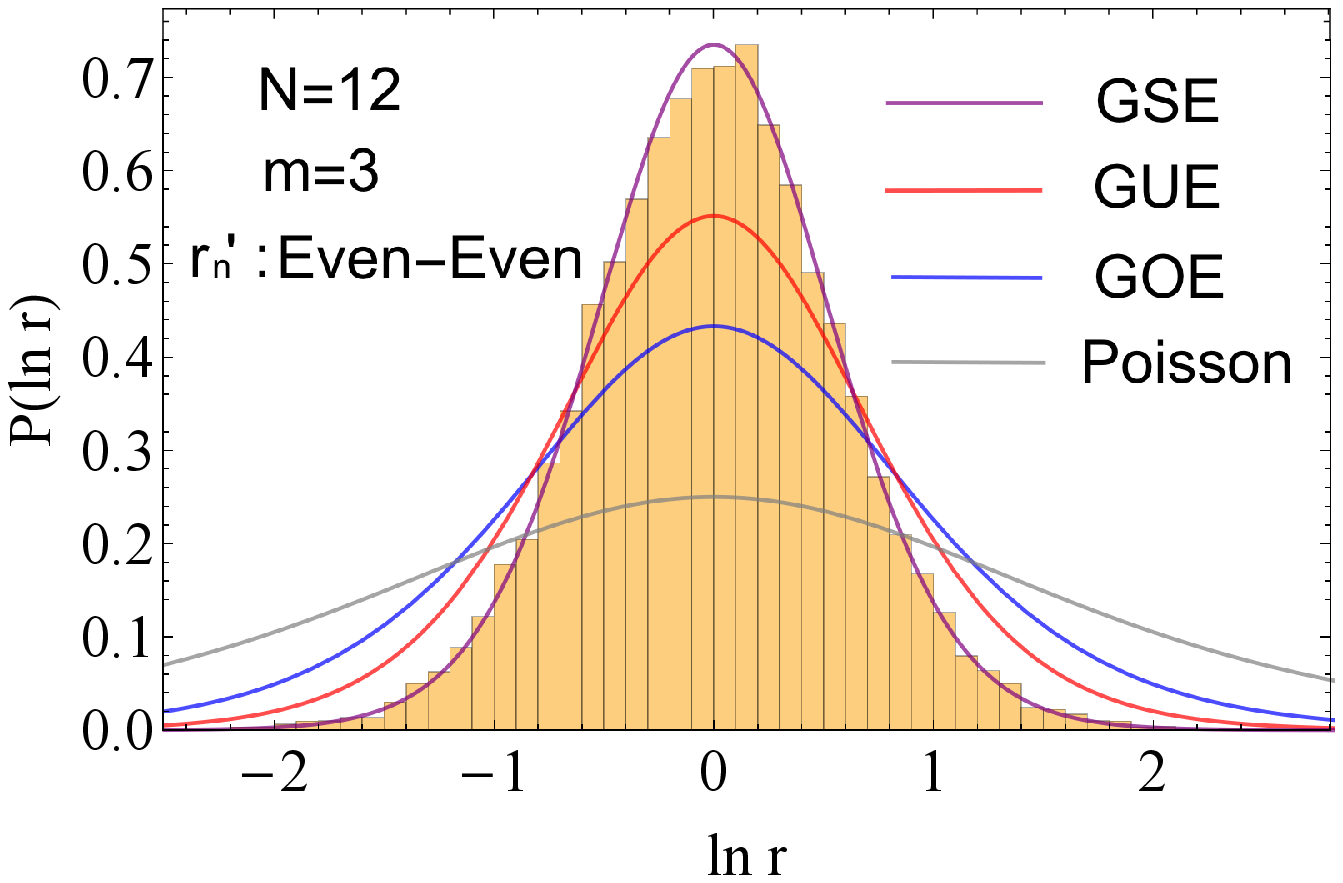}
%\subcaption{Positive tension} 
\end{center}
\end{minipage}
\begin{minipage}{0.49\hsize}
\begin{center}
\includegraphics[width=4.3cm]{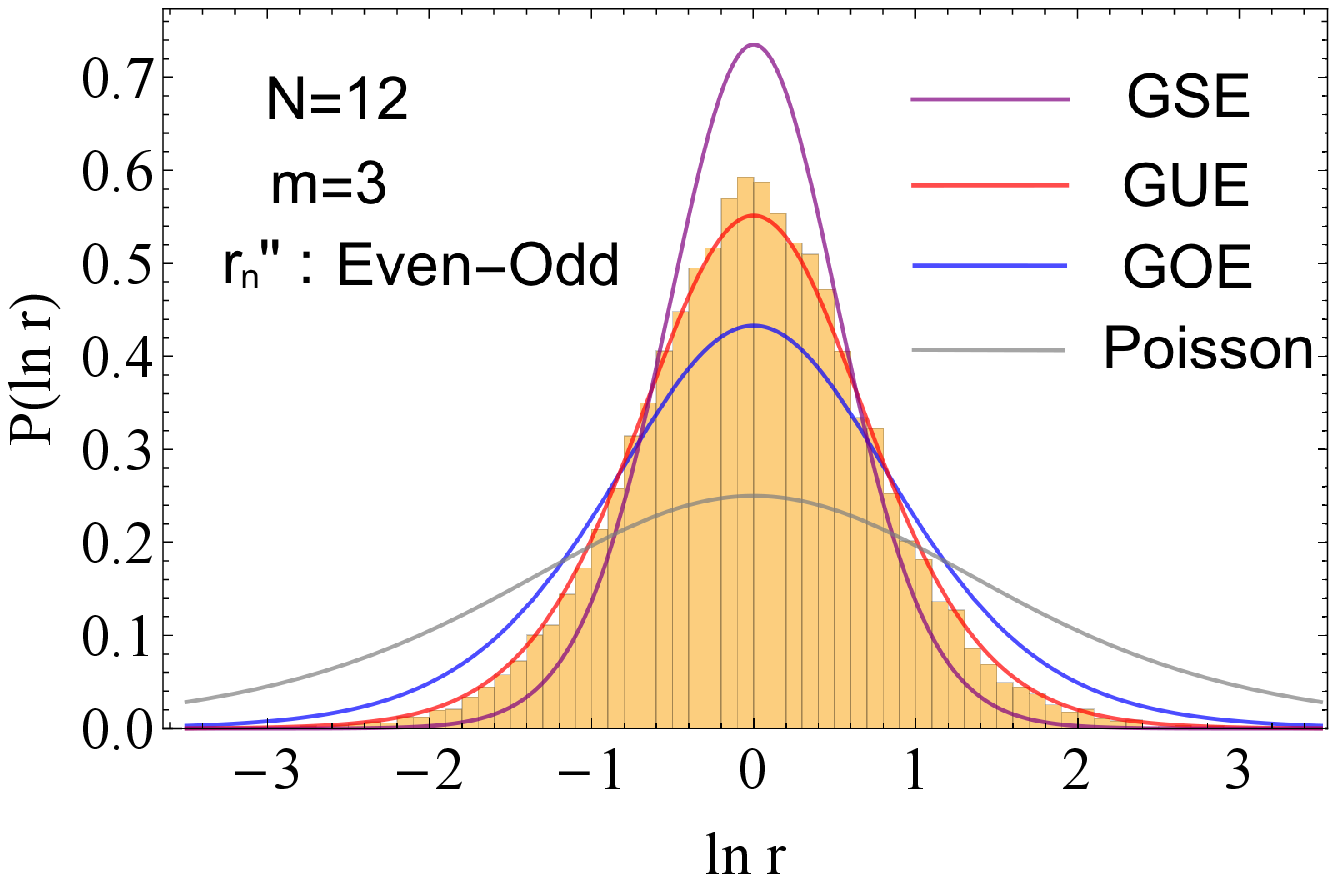}
%\subcaption{Positive tension} 
\end{center}
\end{minipage}
\caption{We plot the probability density $P(\ln r ) = p(r) r$ for some examples. The solid lines are the Wigner surmise given by (\ref{eq:WSpoisson}) and (\ref{eq:WSrandom}).  {\bf Top:} The gap ratio $r_n$ for $m=3$rd band and $m=4$th band in the $N=16, q=4$ deformed SYK model.
We take the average over $1000$ samples.
{\bf Bottom:} We consider the $m=3$rd band in the $N=12,q=4$ deformed SYK model. The left is the gap ratio for even energy levels $r_n'=\f{E_{2n+2}-E_{2n}}{E_{2n}-E_{2n-2}}$. The right is the gap ratio $r_n'' =\f{E_{2n+2}- E_{2n+1}}{E_{2n}-E_{2n-1}}$. We take the average over $2000$ samples for both cases. }
\end{figure}
$m =\f{N}{4}$-th bands in $N=4\ (\text{mod}\ 8)$, the degeneracies are removed at different order of perturbation.
Therefore, $E_{2n}-E_{2n-1}$ is of order $\mu^{-1}$ while $E_{2n+1} - E_{2n}$ is of order $1$.
In this case we expect $E_{2n+2} - E_{2n}$ is determined by the first order perturbation and looks like GSE ensemble when $\mu$ is large.
We study the gap ratio $r_n' = \f{E_{2n+2} - E_{2n}}{E_{2n}- E_{2n-2}}$ numerically and the results show GSE statistics.
On the other hand, the gap ratio $r_n'' =\f{E_{2n+2}- E_{2n+1}}{E_{2n}-E_{2n-1}}$ is also an order one quantity because both of the numerator and the denominator are of order $\mu^{-1}$.
These gap $E_{2n}-E_{2n-1}$ come from (\ref{eq:2ndorder}) that is Hermitian.
Therefore we expect this ratio shows GUE statistics.
We study this gap ratio $r_n''$ numerically and 
%interestingly 
they show GUE statistics as expected.

{\bf{4.Discussion}}
The return amplitude in random matrices is exactly calculated and related to the spectral form factor.
Our numerical study also shows that this relation holds in the SYK model.
Initially the return amplitude decays but at late time first they grows from the dip and then saturates the plateau value.
Because the random matrix behavior is expected to be universal in chaotic systems, we expect that these behaviors are true even in more generic chaotic systems like conformal field theory after a suitable average like a time average.
When we deform the SYK Hamiltonian by a mass term, the return amplitude depends on the choice of initial states.
If we choose the initial product states to be a ground state of the mass term, the return amplitude does not decay.
This deformation prevents the initial state to be scrambled and protects from thermalization.
In gravity side, we can interpret this as the disappearance of the black hole horizon and we have access to the black hole interiors.
If we flip the $m$ spins from the almost ground state one, return amplitude decays and their behaviors are again explained by random matrix theory.
The most interesting case is the $m=\f{N}{4}$ cases in $N=4$ (mod $8$) cases where we see the second dip, ramp and plateau.
In early time the return amplitude behave like GSE statistics with $2$ degeneracies at each level, but at late time it realizes that the degeneracies are actually removed and finally decays to  smaller values.
This serves an example of the return amplitude or the spectral form factor with complicated patterns. 

{\bf{Acknowledgement}}
We thank A.Maloney, S. Harrison and T.Takayanagi for useful discussions.
We also thank R.Namba and D.Yoshida about discussions for numerical calculations.
TN is supported by 
JSPS fellowships and
the Simons Foundation through the It From Qubit collaboration.

%\onecolumngrid
\appendix

\section{A.Explicit Realization of Majorana fermions}
In this appendix we give an explicit representation  of Majorana fermions.
We follow the notation of \cite{Saad:2018bqo}.
We can realize the fermions as the tensor products of the Pauli matrices:
\ba
\psi_{2k-1} &=& \f{1}{\s{2}}Z_1 \cdots Z_{k-1}X_{k}I_{k+1} \cdots I_{N/2}, \notag \\
\psi_{2k} &=& \f{1}{\s{2}}Z_1 \cdots Z_{k-1}Y_{k}I_{k+1} \cdots I_{N/2}. \label{eq:fermions}
\ea
Here 
\be
X_i = 
\begin{pmatrix}
0 & 1 \\
1 & 0
\end{pmatrix}
,\ \ 
Y_i=
\begin{pmatrix}
0 & -i \\
i & 0
\end{pmatrix}
,\ \
Z_i = 
\begin{pmatrix}
1 & 0 \\
0 & -1
\end{pmatrix}
\ee 
are the Pauli operators on $i$-th site  and we omit the symbols $\otimes$ for tensor product in (\ref{eq:fermions}).
Then, the $k$-th spin operator $S_k$ becomes
\be
S_k  = - 2 i \psi_{2k-1}\psi_{2k} = I_1 \dots I_{k-1}Z_k I_{k+1} \dots I_{\f{N}{2}}.
\ee 
This confirms that $S_k$ is the spin operator that measure the eigenvalues of $Z_k$.
In this basis, we can write the $\ket{B_s}$ state as 
\be
\ket{B_s} = \ket{s_1 s_2 \cdots s_{\f{N}{2}}}.
\ee
for the state with $S_k\ket{B_s}= s_k \ket{B_s}$.

We also give the explicit form of  anti unitary symmetry $\mathcal{T}$ and mod $2$ fermion number $(-1)^F$ in this basis.
The fermion number $(-1)^F$ is given by 
\be
(-1)^F =\prod_{k=1}^{\f{N}{2}}S_k =  2^ \f{N}{2} i ^{-\f{N}{2}}\psi_1 \psi_2 \cdots \psi_N. \label{eq:fermionnumber}
\ee
$\mathcal{T}$ depends on the $N$.
When $N/2$ is odd case that corresponds to $N = 2, 6\ (\text{mod}\ 8)$ , 
\be
\mathcal{T} = 2^\f{N}{4}K \psi_1 \psi_3 \cdots\psi_{N-3} \psi_{N-1}, \label{eq:Todd}
\ee
where $K$ is the antiunitary operator that takes the complex conjugate.
When $N/2$ is even case that corresponds to $N = 0, 4\ (\text{mod}\ 8)$ , 
\be
\mathcal{T} = 2^\f{N}{4}K \psi_2 \psi_4 \cdots \psi_{N-2}\psi_{N}.\label{eq:Teven}
\ee
Using this explicit representation, we can show the symmetry property in Table.\ref{table:symmetry}.
Though these realization gives a way to see the symmetry property, they are characterized by topological invariants\cite{PhysRevB.83.075103} and independent from explicit realization.

\section{B. Haar integrals and the derivation of the return amplitude in Random Matrices}
We derive the return amplitude in random matrix theory.
The key observation is that the measure is invariant under unitary conjugation $d(U^{\dagger} H U) e^{-V(U^{\dagger}HU)} = dH e^{-V(H)}$ with $V(H) = \f{L}{2} \Tr H^2$.
Though we choose the potential $V(H)$ of GUE ensemble, we only need the invariance of $V(H)$ under the unitary conjugation.
Then we can represent the GUE ensemble average as $ \braket{f(H)}_{\text{GUE}} = \int dH \int dU e^{-\f{L}{2}\Tr H^2} f(U^{\dagger}H U)$ for any function $f(H)$ where $dU$ is the Haar measure.
For the return amplitude, we get 
\ba
&&\int dH \int dU e^{-V(H)} |\braket{\psi_0|U^{\dagger}e^{-iHt}U|\psi_0}|^2 \notag \\
&=& \int dH \int dU e^{-V(H)}\Tr (e^{-iHt} U \Pi U^{\dagger} e^{iHt} U\Pi U^{\dagger} ) \notag \\
\ea
This integral have four same unitary matrices. 
To evaluate this, we need the following integral:
\ba
&&\int dU_{Haar} U_{i_1j_1}U_{i_2 j_2} U^*_{i_1'j_1'} U^* _{i_2'j_2'} \notag \\
 &=& \f{1}{L^2 -1  } ( \delta_{i_1i_1'}\delta_{i_2i_2'}\delta_{j_1j_1'}\delta_{j_2j_2'}+ \delta_{i_1i_2'}\delta_{i_2i_1'}\delta_{j_1j_2'}\delta_{j_2j_1'}) \notag \\
&&-  \f{1}{L(L^2 -1)  } ( \delta_{i_1i_1'}\delta_{i_2i_2'}\delta_{j_1j_2'}\delta_{j_2j_1'}+ \delta_{i_1i_2'}\delta_{i_2i_1'}\delta_{j_1j_1'}\delta_{j_2j_2'}).\notag \\
\ea
Using this integral, we obtain
\be
\braket{ |\bra{\psi_0}e^{-iHt}\ket{\psi_0}|^2}_{\text{GUE}} =  \f{1}{L(L+1)} (\braket{Z(t)Z(t)^* }_{\text{GUE}}+ L).
\ee
In a similar way, we obtain
\be
\braket{ |\bra{\psi_1}e^{-iHt}\ket{\psi_0}|^2}_{\text{GUE}} =  \f{1}{L^2-1} \Big( L - \f{\braket{Z(t)Z(t)^*}_{\text{GUE}}}{L}\Big),
\ee
for orthogonal $\ket{\psi_0}$ and $\ket{\psi_1}$.

For GSE or GOE cases, the Haar integral for $U\in U(L)$ is replaced by the Haar integral for symplectic groups  $S\in Sp(L/2)$ or orthogonal groups $O \in O(L)$.
The Haar integral for symplectic groups are given by\cite{2006CMaPh.264..773C,1997PhRvB..55.1142A} 
\ba
&&\int dS_{Haar} S_{i_1j_1} S_{i_2j_2}S_{i_1'j_1'}^*S_{i_2'j_2'}^*  \notag \\
&=& \f{L-1}{L(L+1)(L-2)}(\delta_{i_1i_1'}\delta_{i_2i_2'}\delta_{j_1j_1'}\delta_{j_2j_2'} \notag \\
&&+ \delta_{i_1i_2'}\delta_{i_2i_1'}\delta_{j_1j_2'}\delta_{j_2j_1'}  
+ C_{i_1i_2}C_{j_1j_2}C_{i_1'j_1'}C_{j_1'j_2'})  \notag \\
&&-\f{1}{L(L+1)(L-2)}(\delta_{i_1i_1'}\delta_{i_2i_2'}\delta_{j_1j_2'}\delta_{j_2j_1'}   \notag \\
&&+ \delta_{i_1i_2'}\delta_{i_2i_1'}\delta_{j_1j_1'}\delta_{j_2j_2'}  
+ (\delta_{i_1i_1'}\delta_{i_2i_2'} - \delta_{i_1i_2'}\delta_{i_2i_1'}) C_{j_1j_2}C_{j_1'j_2'} \notag \\
&&+ C_{i_1i_2}C_{i_1'i_2'} (\delta_{j_1j_1'}\delta_{j_2j_2'}-\delta_{j_1j_2'}\delta_{j_2j_1'}) ),
\ea
where $C_{ij}$ is the antisymmetric invariant.
This coupling gives the inner product between the Kramers' pairs.
Because this coupling is antisymmetric, the diagonal part $v_i C_{ij}v_j$ for $\ket{\psi_{0}} = \sum_i v_i \ket{e_i} $ vanishes and they can be ignored in the calculation of the return amplitude.
Using this integral, we obtain
\be
\braket{ |\bra{\psi_0}e^{-iHt}\ket{\psi_0}|^2}_{\text{GSE}} =  \f{1}{L(L+1)} (\braket{Z(t)Z(t)^* }_{\text{GSE}}+ L),
\ee
which is the same expression with the GUE case though the Haar integrals themselves are different.

The Haar integral for orthogonal groups are given by\cite{2006CMaPh.264..773C,1997PhRvB..55.1142A} 
\ba
&&\int dO_{Haar} O_{i_1j_1} O_{i_2j_2}O_{i_1'j_1'}^*O_{i_2'j_2'}^*  \notag \\
&=& \f{L+1}{L(L-1)(L+2)}(\delta_{i_1i_1'}\delta_{i_2i_2'}\delta_{j_1j_1'}\delta_{j_2j_2'}
\notag \\
&&+ \delta_{i_1i_2'}\delta_{i_2i_1'}\delta_{j_1j_2'}\delta_{j_2j_1'}  + C_{i_1i_2}C_{j_1j_2}C_{i_1'j_1'}C_{j_1'j_2'}) \notag \\
&&-\f{1}{L(L-1)(L+2)}(\delta_{i_1i_1'}\delta_{i_2i_2'}\delta_{j_1j_2'}\delta_{j_2j_1'} \notag \\
&&+ \delta_{i_1i_2'}\delta_{i_2i_1'}\delta_{j_1j_1'}\delta_{j_2j_2'} 
+ (\delta_{i_1i_1'}\delta_{i_2i_2'} + \delta_{i_1i_2'}\delta_{i_2i_1'}) C_{j_1j_2}C_{j_1'j_2'} \notag \\
&&+ C_{i_1i_2}C_{i_1'i_2'} (\delta_{j_1j_1'}\delta_{j_2j_2'}+\delta_{j_1j_2'}\delta_{j_2j_1'}) ),
\ea
where $C_{ij}$ is the symmetric coupling and we can choose a basis with $C_{ij} = \delta_{ij}$.
Unlike the case of symplectic groups, the diagonal part $v_i C_{ij}v_j$ for $\ket{\psi_{0}} = \sum_i v_i \ket{e_i} $ does not vanish  and the return amplitude depends on states.

\section{C. Symmetry analysis of the Deformed SYK Hamiltonian  at 1st order perturbation}
In this section we study the symmetry class $H_m$ in detail that is imposed by $\mathcal{T}$ symmetry.
Because $\mathcal{T}$ relates the $m$-th band and the $(\f{N}{2} -m)$-th band, it is sufficient to consider the the symmetry constraint on the following submatrix:
\be
H^{m,\f{N}{2}-m} = 
\begin{pmatrix}
 H_m & S_m \\
 S_m^{\dagger} & H_{\f{N}{2}-m}
\end{pmatrix},
\ee
where $H_m = P_{m}H_{SYK}P_m$ 
%is Hermitian 
and $S_{m} = P_m H_{SYK}P_{\f{N}{2}-m}$.
% is the complex matrix.
Using the symmetry of the SYK Hamiltonian $(-1)^FH_{SYK}(-1)^F = H_{SYK}$, 
we find $P_mH_{SYK}P_{\f{N}{2}-m}=(-1)^{\f{N}{2}}P_mH_{SYK}P_{\f{N}{2}-m}$.
This means $S_m = 0$ for $N=2,6\ (\text{mod}\ 8 )$.
For $N=0,2 \ (\text{mod}\ 8)$ cases,  $\mathcal{T}^2= 1$. 
In these cases, we can choose the basis with
$
\mathcal{T} = \begin{pmatrix}
 0 & \mathbb{I}_{d_m} \\
 \mathbb{I}_{d_m}& 0
\end{pmatrix} K
$
where $\mathbb{I}_{d_m}$ is the identity matrix of rank $d_m$ and $K$ is the complex conjugate operator.
The invariance under this $\mathcal{T}$ imposes the condition $H_{\f{N}{2} - m}^* = H_m$ and $S_m^T = S_m$.
For $N=0\ (\text{mod}\ 8)$ this means that $H^{m,\f{N}{2}-m}$ is conjugate to real symmetric matrix though $H^{m,\f{N}{2}-m}$  is expressed in unusual basis in which the reality is not manifest.
To see the reality manifestly, it is convenient to change the basis in the following way:
\ba
&&
 \begin{pmatrix}
 H_m & S_{m} \\
 S_{m}^* & H_{m}^*
\end{pmatrix} \notag \\
&&\to
\f{1}{\s{2}}\begin{pmatrix}
 \mathbb{I}_{d_m} & \mathbb{I}_{d_m} \\
 i \mathbb{I}_{d_m}& -i\mathbb{I}_{d_m}
\end{pmatrix}
 \begin{pmatrix}
 H_m & S_{m} \\
 S_{m}^* & H_{m}^*
\end{pmatrix}
\f{1}{\s{2}}\begin{pmatrix}
 \mathbb{I}_{d_m} & -i\mathbb{I}_{d_m} \\
 \mathbb{I}_{d_m} & i\mathbb{I}_{d_m}
\end{pmatrix} \notag \\
%&=& \f{1}{2} 
%\begin{pmatrix}
%H_m + H_m^* + S_m + S_m^* & - i (H_m-H_m^*)+ i (S_m - S_m^*) \\
% i (H_m-H_m^*)+ i (S_m - S_m^*) & H_m + H_m^* - S_m - S_m^*
%\end{pmatrix}
%\notag  \\
&=& 
\begin{pmatrix}
\text{Re} H_m +\text{Re} S_m & \text{Im} H_m -  \text{Im}S_m \\
 -\text{Im} H_m -  \text{Im}S_m &  \text{Re} H_m -\text{Re} S_m
\end{pmatrix}.
\ea
This takes the form of the real symmetric matrix under the condition $S_m^T = S_m$ and $H_m^{\dagger} = H_m$. 
Though $\mathcal{T}$ relate $H_m$ and $H_{\f{N}{2}-m}$, it does not impose any constraint on $H_m$ itself.
Therefore, for generic matrix $H^{m,\f{N}{2}-m}$, $H_m$ belongs to GUE ensemble.

For $N = 4,6 \ (\text{mod}\ 8)$, $\mathcal{T}^2 = -1$. 
In these cases, we can choose the basis with 
$
\mathcal{T} = \begin{pmatrix}
 0 & -\mathbb{I}_{d_m} \\
 \mathbb{I}_{d_m}& 0
\end{pmatrix} K.
$
In the same manner, we obtain $H_m = H_{\f{N}{2}-m}^*$ and $S_m^T = -S_m$.
For $N=4\ (\text{mod}\ 8 )$ this means that $H^{m,\f{N}{2}-m}$ is a quaternion Hermitian.
We can write as $H_m = H  + iA_z$  and $S_{m} = i A_x + A_y$ where $H$ is real symmetric and $A_x,A_y,A_z$ is real antisymmetric.
Then, $H ^{m,\f{N}{2}-m}= H \otimes \mathbb{I}_2 +\sum_{\alpha}i A_{\alpha}\otimes\sigma_\alpha$ gives a usual realization of quaternion Hermitian matrices.
$H_m = H + i A_z$ spans generic Hermitian matrices and the ensemble is GUE.
Again though $\mathcal{T}$ relate $H_m$ and $H_{\f{N}{2}-m}$, $\mathcal{T}$ does not impose any condition on $H_{m}$ itself.

\section{D. Evolution Amplitude in the SYK model}
\begin{figure}[htbp]
\begin{center}
\includegraphics[width=\hsize]{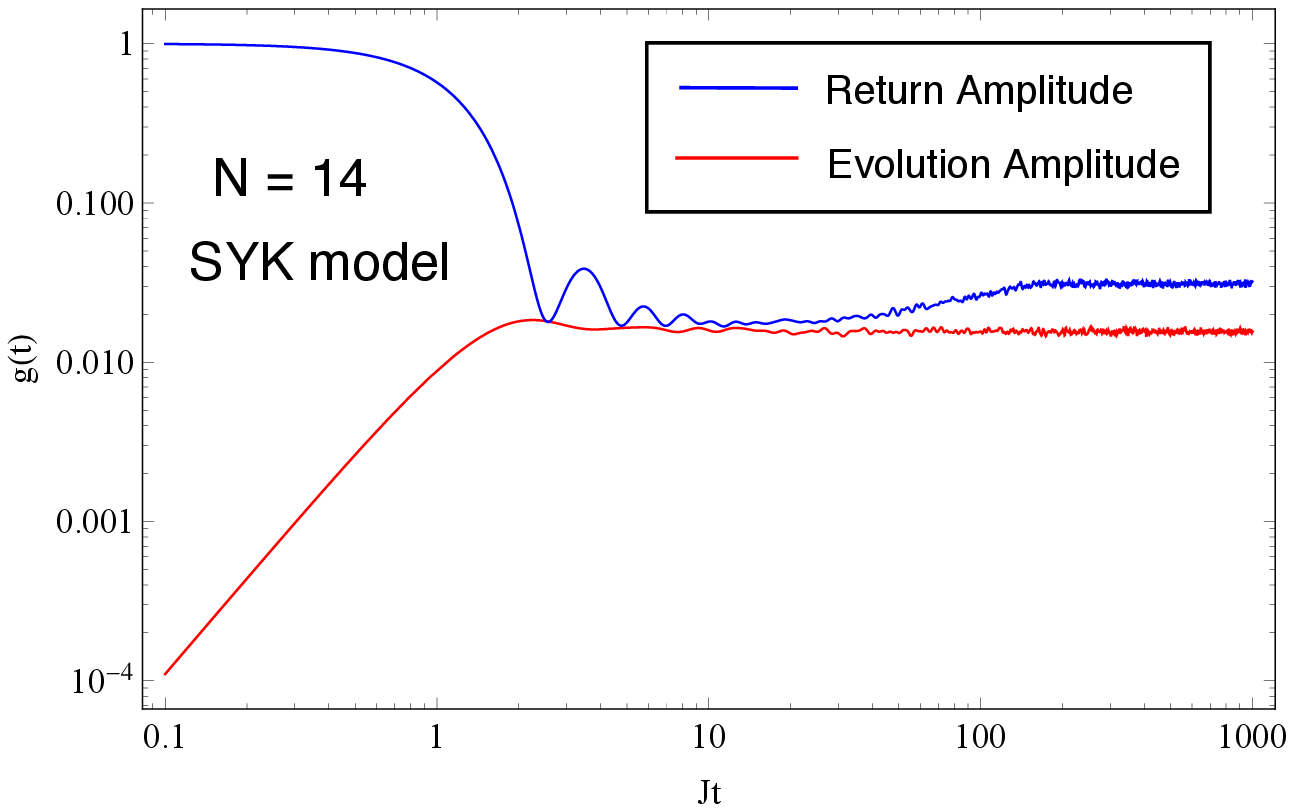}
\includegraphics[width=\hsize]{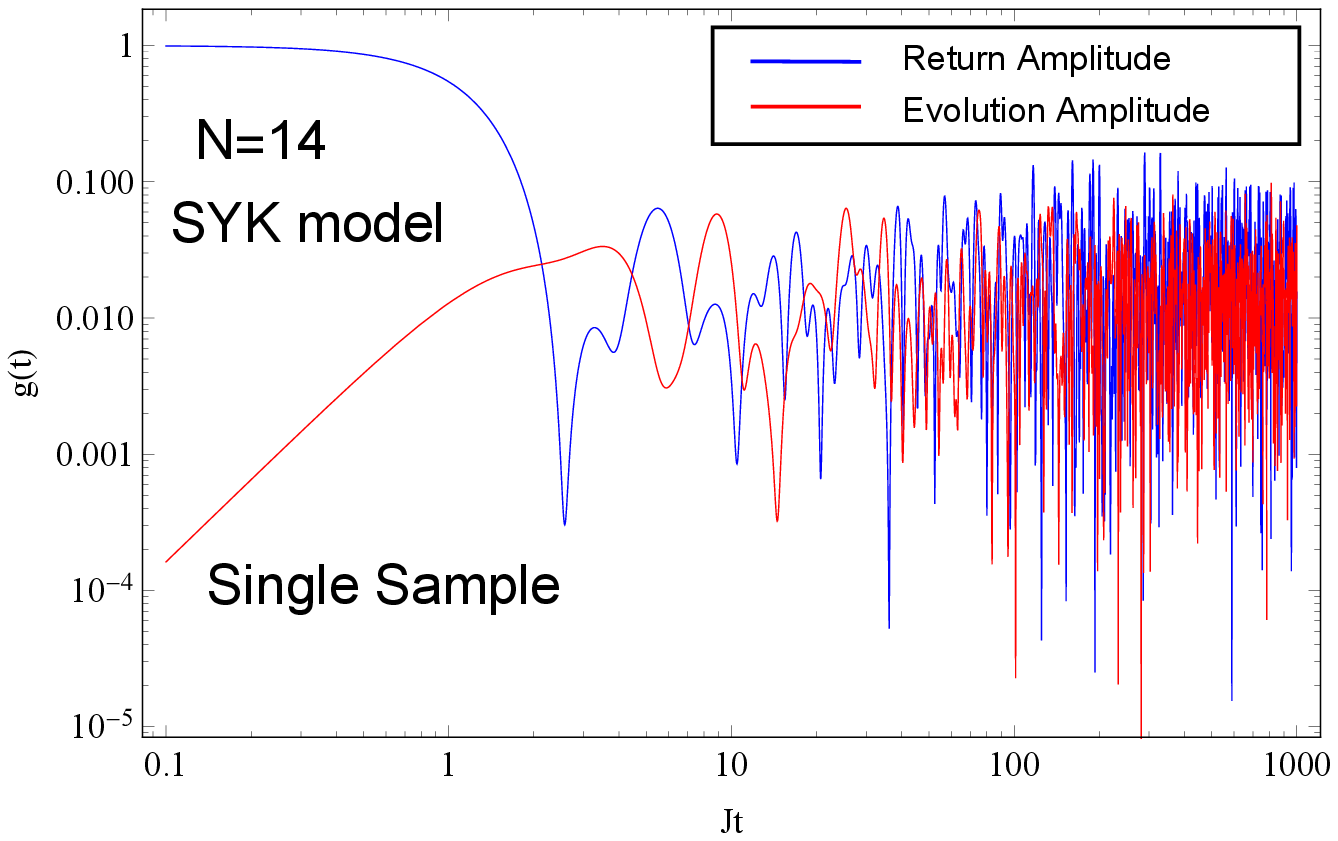}
\caption{These are the numerical plots of the return amplitude and the evolution amplitude in the $N=14, q=4 $ SYK model.
As an example of the return amplitude, we consider $|\bra{B_{\uparrow\uparrow\uparrow\uparrow\uparrow\uparrow\uparrow}}e^{-iH_{SYK}t}\ket{B_{\uparrow\uparrow\uparrow\uparrow\uparrow\uparrow\uparrow}}|^2$, and  as an example of the evolution amplitude we consider  $|\bra{B_{\downarrow\downarrow\uparrow\uparrow\uparrow\uparrow\uparrow}}e^{-iH_{SYK}t}\ket{B_{\uparrow\uparrow\uparrow\uparrow\uparrow\uparrow\uparrow}}|^2$.
{\bf Top:} We take the average over $1500$ samples for both of the return amplitude and the evolution amplitude.
{\bf Bottom:} The return amplitude and the evolution amplitude with single sample.
}
\label{fig:RAvsEAav2}
\end{center}
\end{figure}

In this section we consider the overlap between time evolved state $e^{-iHt}\ket{\psi_0}$ and states $\ket{\psi_1}$  that is orthogonal to $\ket{\psi_0}$, which we consider in random matrices in (\ref{eq:EAdef}), in the SYK model.
Especially, we consider the amplitude between $\ket{B_s}$ with different spins:
\be
|\bra{B_{s'}}e^{-iH_{SYK}t}\ket{B_s}|^2.
\ee
Here we call this the {\it evolution amplitude}.
In figure.\ref{fig:RAvsEAav2}, we compare the return amplitude and the evolution amplitude numerically in the SYK model.
%in $N=14, q = 4$ SYK model.
After the ensemble average, they show that the plateau value in the return amplitude is clearly larger than the evolution amplitude. 
Even in single sample, the return amplitude looks to oscillate around the larger average value than the the evolution amplitude.

\section{E. Time Average of single sample in the SYK model}
\begin{figure}[htbp]
\begin{center}
\includegraphics[width=\hsize]{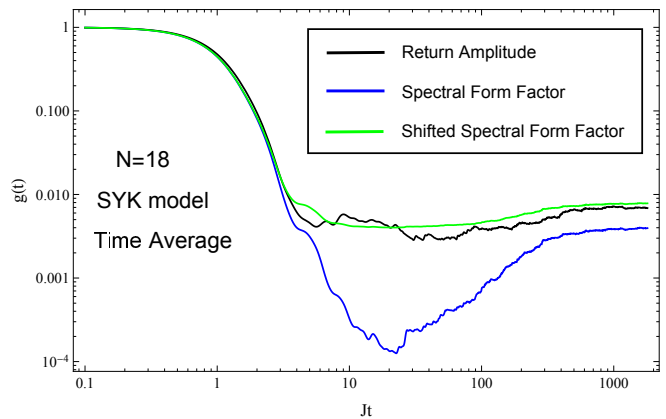}
\caption{This is the plot of the return amplitude in the $N=18, q=4$ SYK model with the time average.
We choose $\ket{B_{\uparrow\uparrow\uparrow\uparrow\uparrow\uparrow\uparrow\uparrow\uparrow}}$ as the initial state.
We can see the slope, dip, ramp and plateau even in the time average cases.
 As we mentioned, the plateau value of the time averaged return amplitude have a small deviation from the shifted spectral form factor.
} 
\label{fig:Tave2}
\end{center}
\end{figure}

The spectral form factor is not self averaging\cite{1997PhRvL..78.2280P}, and we need some averages to get smooth behaviors.
In the SYK model, we take the ensemble average over the coupling $J_{a_1\cdots a_q}$.
In this appendix, we consider the time average of the return amplitude in a single sample in the SYK model, which is another kind of average.
First we consider the infinite time average , which gives the averaged plateau value, in general quantum systems without degeneracy.
By decomposing the state $\ket{\psi_0} = \sum_i c_i \ket{E_i}$, we obtain
\ba
&&\lim_{T\to\infty}\f{1}{T} \int _0^{T} |\bra{\psi_0}e^{-iHt}\ket{\psi_0}|^2  \notag \\
&=& \lim_{T\to\infty}\f{1}{T} \int _0^{T} dt \sum_{i,j} c_i^*c_ic_j^*c_j e^{-i(E_i-E_j)t} \notag \\
&=& \sum_i |c_i|^4.\label{eq:timeaveinf}
\ea
Therefore the time average is not exactly the same with the plateau value in ensemble average (\ref{eq:RAandSFF}).
If we take the average of (\ref{eq:timeaveinf}) over states , it becomes the late time value $2/(L+1)$  in the ensemble average.
In the SYK model, we consider the following time average:
\be
g_p^{time}(t;\beta) = \int_{\f{1}{2}t}^{\f{3}{2}t} |\bra{B_s}e^{-iH_{SYK}t' - \beta H_{SYK}}\ket{B_s}|^2 dt'.
\ee
Here we take the time average between $\f{1}{2}t < t' < \f{3}{2}t$ around time $t$, which is taken in\cite{Balasubramanian:2016ids}.
We show the plots of the return amplitude and the spectral form factor with time average in figure.\ref{fig:Tave2}.
Both of the return amplitude and the spectral form factor show the slope, dip, ramp and plateau behavior.
This motivate us to expect that the return amplitude in more generic quantum systems like chaotic CFTs shows these structure after the time average.
We also compare the shifted spectral form factor (\ref{eq:timeaveinf}) where the ensemble average of the spectral form factor is replaced by the time average.
As we pointed out, their late time value is not exactly the same, but the behavior shows good agreement on each time.

\bibliography{traversablebunken}

\providecommand{\href}[2]{#2}\begingroup\raggedright\begin{thebibliography}{10}

\bibitem{PhysRevLett.70.3339}
S.~Sachdev and J.~Ye \href{http://dx.doi.org/10.1103/PhysRevLett.70.3339}{{\em
  Phys. Rev. Lett.} {\bfseries 70} (May, 1993) 3339--3342}.

\bibitem{KitaevTalk}
A.~Kitaev {\em A simple model of quantum holography",talk at KITP strings
  seminar and Entanglement 2015 program} (2015) .

\bibitem{Maldacena:2016hyu}
J.~Maldacena and D.~Stanford
  \href{http://dx.doi.org/10.1103/PhysRevD.94.106002}{{\em Phys. Rev.}
  {\bfseries D94} no.~10, (2016) 106002},
\href{http://arxiv.org/abs/1604.07818}{{\ttfamily arXiv:1604.07818 [hep-th]}}.
%%CITATION = ARXIV:1604.07818;%%.

\bibitem{Kitaev:2017awl}
A.~Kitaev and S.~J. Suh \href{http://dx.doi.org/10.1007/JHEP05(2018)183}{{\em
  JHEP} {\bfseries 05} (2018) 183},
\href{http://arxiv.org/abs/1711.08467}{{\ttfamily arXiv:1711.08467 [hep-th]}}.
%%CITATION = ARXIV:1711.08467;%%.

\bibitem{Maldacena:2015waa}
J.~Maldacena, S.~H. Shenker, and D.~Stanford
  \href{http://dx.doi.org/10.1007/JHEP08(2016)106}{{\em JHEP} {\bfseries 08}
  (2016) 106},
\href{http://arxiv.org/abs/1503.01409}{{\ttfamily arXiv:1503.01409 [hep-th]}}.
%%CITATION = ARXIV:1503.01409;%%.

\bibitem{PhysRevB.95.115150}
Y.-Z. You, A.~W.~W. Ludwig, and C.~Xu
  \href{http://dx.doi.org/10.1103/PhysRevB.95.115150}{{\em Phys. Rev. B}
  {\bfseries 95} (Mar, 2017) 115150}.

\bibitem{Cotler:2016fpe}
J.~S. Cotler, G.~Gur-Ari, M.~Hanada, J.~Polchinski, P.~Saad, S.~H. Shenker,
  D.~Stanford, A.~Streicher, and M.~Tezuka
  \href{http://dx.doi.org/10.1007/JHEP09(2018)002,
  10.1007/JHEP05(2017)118}{{\em JHEP} {\bfseries 05} (2017) 118},
  \href{http://arxiv.org/abs/1611.04650}{{\ttfamily arXiv:1611.04650
  [hep-th]}}.
[Erratum: JHEP09,002(2018)].
%%CITATION = ARXIV:1611.04650;%%.

\bibitem{JACKIW1985343}
R.~Jackiw
  \href{http://dx.doi.org/https://doi.org/10.1016/0550-3213(85)90448-1}{{\em
  Nuclear Physics B} {\bfseries 252} (1985) 343 -- 356}.

\bibitem{Teitelboim:1983ux}
C.~Teitelboim
\href{http://dx.doi.org/10.1016/0370-2693(83)90012-6}{{\em Phys. Lett.}
  {\bfseries 126B} (1983) 41--45}.
%%CITATION = PHLTA,126B,41;%%.

\bibitem{Almheiri:2014cka}
A.~Almheiri and J.~Polchinski
  \href{http://dx.doi.org/10.1007/JHEP11(2015)014}{{\em JHEP} {\bfseries 11}
  (2015) 014},
\href{http://arxiv.org/abs/1402.6334}{{\ttfamily arXiv:1402.6334 [hep-th]}}.
%%CITATION = ARXIV:1402.6334;%%.

\bibitem{Maldacena:2016upp}
J.~Maldacena, D.~Stanford, and Z.~Yang
  \href{http://dx.doi.org/10.1093/ptep/ptw124}{{\em PTEP} {\bfseries 2016}
  no.~12, (2016) 12C104},
\href{http://arxiv.org/abs/1606.01857}{{\ttfamily arXiv:1606.01857 [hep-th]}}.
%%CITATION = ARXIV:1606.01857;%%.

\bibitem{Kourkoulou:2017zaj}
I.~Kourkoulou and J.~Maldacena
\href{http://arxiv.org/abs/1707.02325}{{\ttfamily arXiv:1707.02325 [hep-th]}}.
%%CITATION = ARXIV:1707.02325;%%.

\bibitem{Krishnan:2017txw}
C.~Krishnan and K.~V.~P. Kumar
  \href{http://dx.doi.org/10.1007/JHEP10(2017)099}{{\em JHEP} {\bfseries 10}
  (2017) 099},
\href{http://arxiv.org/abs/1706.05364}{{\ttfamily arXiv:1706.05364 [hep-th]}}.
%%CITATION = ARXIV:1706.05364;%%.

\bibitem{Hunter-Jones:2017raw}
N.~Hunter-Jones, J.~Liu, and Y.~Zhou
  \href{http://dx.doi.org/10.1007/JHEP02(2018)142}{{\em JHEP} {\bfseries 02}
  (2018) 142},
\href{http://arxiv.org/abs/1710.03012}{{\ttfamily arXiv:1710.03012 [hep-th]}}.
%%CITATION = ARXIV:1710.03012;%%.

\bibitem{Dhar:2018pii}
A.~Dhar, A.~Gaikwad, L.~K. Joshi, G.~Mandal, and S.~R. Wadia
\href{http://arxiv.org/abs/1812.03979}{{\ttfamily arXiv:1812.03979 [hep-th]}}.
%%CITATION = ARXIV:1812.03979;%%.

\bibitem{Bhattacharya:2018fkq}
R.~Bhattacharya, D.~P. Jatkar, and N.~Sorokhaibam
\href{http://arxiv.org/abs/1811.06006}{{\ttfamily arXiv:1811.06006 [hep-th]}}.
%%CITATION = ARXIV:1811.06006;%%.

\bibitem{Numasawa:2016emc}
T.~Numasawa, N.~Shiba, T.~Takayanagi, and K.~Watanabe
  \href{http://dx.doi.org/10.1007/JHEP08(2016)077}{{\em JHEP} {\bfseries 08}
  (2016) 077},
\href{http://arxiv.org/abs/1604.01772}{{\ttfamily arXiv:1604.01772 [hep-th]}}.
%%CITATION = ARXIV:1604.01772;%%.

\bibitem{Cardy:2014rqa}
J.~Cardy \href{http://dx.doi.org/10.1103/PhysRevLett.112.220401}{{\em Phys.
  Rev. Lett.} {\bfseries 112} (2014) 220401},
\href{http://arxiv.org/abs/1403.3040}{{\ttfamily arXiv:1403.3040
  [cond-mat.stat-mech]}}.
%%CITATION = ARXIV:1403.3040;%%.

\bibitem{Cardy:2016lei}
J.~Cardy \href{http://dx.doi.org/10.1088/1751-8113/49/41/415401}{{\em J. Phys.}
  {\bfseries A49} no.~41, (2016) 415401},
\href{http://arxiv.org/abs/1603.08267}{{\ttfamily arXiv:1603.08267
  [cond-mat.stat-mech]}}.
%%CITATION = ARXIV:1603.08267;%%.

\bibitem{Brustein:2018fkr}
R.~Brustein and Y.~Zigdon
  \href{http://dx.doi.org/10.1103/PhysRevD.98.066013}{{\em Phys. Rev.}
  {\bfseries D98} no.~6, (2018) 066013},
\href{http://arxiv.org/abs/1804.09017}{{\ttfamily arXiv:1804.09017 [hep-th]}}.
%%CITATION = ARXIV:1804.09017;%%.

\bibitem{PhysRevB.83.075103}
L.~Fidkowski and A.~Kitaev
  \href{http://dx.doi.org/10.1103/PhysRevB.83.075103}{{\em Phys. Rev. B}
  {\bfseries 83} (Feb, 2011) 075103}.

\bibitem{Kanazawa:2017dpd}
T.~Kanazawa and T.~Wettig \href{http://dx.doi.org/10.1007/JHEP09(2017)050}{{\em
  JHEP} {\bfseries 09} (2017) 050},
\href{http://arxiv.org/abs/1706.03044}{{\ttfamily arXiv:1706.03044 [hep-th]}}.
%%CITATION = ARXIV:1706.03044;%%.

\bibitem{Balasubramanian:2016ids}
V.~Balasubramanian, B.~Craps, B.~Czech, and G.~S^^c3^^a1rosi
  \href{http://dx.doi.org/10.1007/JHEP03(2017)154}{{\em JHEP} {\bfseries 03}
  (2017) 154},
\href{http://arxiv.org/abs/1612.04334}{{\ttfamily arXiv:1612.04334 [hep-th]}}.
%%CITATION = ARXIV:1612.04334;%%.

\bibitem{Benjamin:2018kre}
N.~Benjamin, E.~Dyer, A.~L. Fitzpatrick, and Y.~Xin
\href{http://arxiv.org/abs/1812.07579}{{\ttfamily arXiv:1812.07579 [hep-th]}}.
%%CITATION = ARXIV:1812.07579;%%.

\bibitem{Chenu:2018spm}
A.~Chenu, J.~Molina-Vilaplana, and A.~Del~Campo
\href{http://arxiv.org/abs/1804.09188}{{\ttfamily arXiv:1804.09188
  [quant-ph]}}.
%%CITATION = ARXIV:1804.09188;%%.

\bibitem{delCampo:2017bzr}
A.~del Campo, J.~Molina-Vilaplana, and J.~Sonner
  \href{http://dx.doi.org/10.1103/PhysRevD.95.126008}{{\em Phys. Rev.}
  {\bfseries D95} no.~12, (2017) 126008},
\href{http://arxiv.org/abs/1702.04350}{{\ttfamily arXiv:1702.04350 [hep-th]}}.
%%CITATION = ARXIV:1702.04350;%%.

\bibitem{2006CMaPh.264..773C}
B.~{Collins} and P.~{{\'S}niady}
  \href{http://dx.doi.org/10.1007/s00220-006-1554-3}{{\em Communications in
  Mathematical Physics} {\bfseries 264} (June, 2006) 773--795},
  \href{http://arxiv.org/abs/math-ph/0402073}{{\ttfamily arXiv:math-ph/0402073
  [math-ph]}}.

\bibitem{1997PhRvB..55.1142A}
A.~{Altland} and M.~R. {Zirnbauer}
  \href{http://dx.doi.org/10.1103/PhysRevB.55.1142}{{\em Physical Review B}
  {\bfseries 55} (Jan., 1997) 1142--1161},
  \href{http://arxiv.org/abs/cond-mat/9602137}{{\ttfamily
  arXiv:cond-mat/9602137 [cond-mat]}}.

\bibitem{2018arXiv180707577S}
M.~{Schiulaz}, E.~J. {Torres-Herrera}, and L.~F. {Santos} {\em arXiv e-prints}
  (July, 2018) arXiv:1807.07577,
  \href{http://arxiv.org/abs/1807.07577}{{\ttfamily arXiv:1807.07577
  [cond-mat.stat-mech]}}.

\bibitem{Saad:2018bqo}
P.~Saad, S.~H. Shenker, and D.~Stanford
\href{http://arxiv.org/abs/1806.06840}{{\ttfamily arXiv:1806.06840 [hep-th]}}.
%%CITATION = ARXIV:1806.06840;%%.

\bibitem{Cotler:2017jue}
J.~Cotler, N.~Hunter-Jones, J.~Liu, and B.~Yoshida
  \href{http://dx.doi.org/10.1007/JHEP11(2017)048}{{\em JHEP} {\bfseries 11}
  (2017) 048},
\href{http://arxiv.org/abs/1706.05400}{{\ttfamily arXiv:1706.05400 [hep-th]}}.
%%CITATION = ARXIV:1706.05400;%%.

\bibitem{1997PhRvL..78.2280P}
R.~E. {Prange} \href{http://dx.doi.org/10.1103/PhysRevLett.78.2280}{{\em \prl}
  {\bfseries 78} (Mar., 1997) 2280--2283},
  \href{http://arxiv.org/abs/chao-dyn/9606010}{{\ttfamily
  arXiv:chao-dyn/9606010 [nlin.CD]}}.

\bibitem{Eberlein:2017wah}
A.~Eberlein, V.~Kasper, S.~Sachdev, and J.~Steinberg
  \href{http://dx.doi.org/10.1103/PhysRevB.96.205123}{{\em Phys. Rev.}
  {\bfseries B96} no.~20, (2017) 205123},
\href{http://arxiv.org/abs/1706.07803}{{\ttfamily arXiv:1706.07803
  [cond-mat.str-el]}}.
%%CITATION = ARXIV:1706.07803;%%.

\bibitem{Garcia-Garcia:2017bkg}
A.~M. Garc^^c3^^ada-Garc^^c3^^ada, B.~Loureiro, A.~Romero-Berm^^c3^^badez, and
  M.~Tezuka \href{http://dx.doi.org/10.1103/PhysRevLett.120.241603}{{\em Phys.
  Rev. Lett.} {\bfseries 120} no.~24, (2018) 241603},
\href{http://arxiv.org/abs/1707.02197}{{\ttfamily arXiv:1707.02197 [hep-th]}}.
%%CITATION = ARXIV:1707.02197;%%.

\bibitem{Nosaka:2018iat}
T.~Nosaka, D.~Rosa, and J.~Yoon
  \href{http://dx.doi.org/10.1007/JHEP09(2018)041}{{\em JHEP} {\bfseries 09}
  (2018) 041},
\href{http://arxiv.org/abs/1804.09934}{{\ttfamily arXiv:1804.09934 [hep-th]}}.
%%CITATION = ARXIV:1804.09934;%%.

\bibitem{messiah1999quantum}
A.~Messiah, {\em Quantum Mechanics}.
\newblock Dover books on physics. Dover Publications, 1999.
\newblock \url{https://books.google.ca/books?id=mwssSDXzkNcC}.

\bibitem{2015PhRvB..91h1103L}
D.~J. {Luitz}, N.~{Laflorencie}, and F.~{Alet}
  \href{http://dx.doi.org/10.1103/PhysRevB.91.081103}{{\em \prb} {\bfseries 91}
  no.~8, (Feb., 2015) 081103}, \href{http://arxiv.org/abs/1411.0660}{{\ttfamily
  arXiv:1411.0660 [cond-mat.dis-nn]}}.

\bibitem{2007PhRvB..75o5111O}
V.~{Oganesyan} and D.~A. {Huse}
  \href{http://dx.doi.org/10.1103/PhysRevB.75.155111}{{\em \prb} {\bfseries 75}
  no.~15, (Apr., 2007) 155111},
  \href{http://arxiv.org/abs/cond-mat/0610854}{{\ttfamily cond-mat/0610854}}.

\bibitem{2016PhRvB..94n4201B}
C.~L. {Bertrand} and A.~M. {Garc{\'{\i}}a-Garc{\'{\i}}a}
  \href{http://dx.doi.org/10.1103/PhysRevB.94.144201}{{\em \prb} {\bfseries 94}
  no.~14, (Oct., 2016) 144201},
  \href{http://arxiv.org/abs/1606.08419}{{\ttfamily arXiv:1606.08419
  [cond-mat.dis-nn]}}.

\bibitem{2013PhRvL.110h4101A}
Y.~Y. {Atas}, E.~{Bogomolny}, O.~{Giraud}, and G.~{Roux}
  \href{http://dx.doi.org/10.1103/PhysRevLett.110.084101}{{\em Physical Review
  Letters} {\bfseries 110} no.~8, (Feb., 2013) 084101},
  \href{http://arxiv.org/abs/1212.5611}{{\ttfamily arXiv:1212.5611 [math-ph]}}.

\end{thebibliography}\endgroup

\end{document}